\documentclass[10pt,sigconf,letterpaper,nonacm]{acmart}

\def\Snospace~{\S{}}

\usepackage{amsmath}
\usepackage{epsfig,endnotes}
\usepackage{epstopdf}
\usepackage{xspace}
\usepackage{subcaption}
\usepackage{color}
\usepackage{paralist}
\usepackage{threeparttable}
\usepackage{fancyvrb}
\usepackage{relsize}
\usepackage{bbm}
\usepackage{booktabs}
\usepackage[noend]{algpseudocode}
\usepackage{soul}
\usepackage{subfloat}
\usepackage{multirow}
\usepackage{balance}
\usepackage{flushend}
\usepackage{makecell}
\usepackage{hhline}
\usepackage{url}
\usepackage{graphicx}
\usepackage[all]{nowidow}
\usepackage{pifont}
\usepackage{epigraph}
\usepackage{etoolbox}
\usepackage[final]{pdfpages}
\usepackage[normalem]{ulem}
\usepackage[linesnumbered,ruled,vlined,noend]{algorithm2e}

\usepackage[compact]{titlesec}
\SetAlCapNameFnt{\small}
\SetAlCapFnt{\small}
\def\Snospace~{\S{}}

\newcommand{\ie}[0]{\textit{i.e.,}\xspace}
\newcommand{\eg}[0]{\textit{e.g.,}\xspace}

\SetCommentSty{mycommfont}

\usepackage{tikz}

\usepackage{ifthen}
\usepackage{xcolor}

\newcommand{\exclude}[1]{}
\newcommand{\showComments}{yes}
\newcommand{\note}[2]{
\ifthenelse{\equal{\showComments}{yes}}{\textcolor{#1}{#2}}{} }

\newcommand{\tightsection}[1]{\section{#1}\vspace{-0.05in}}
\newcommand{\tightsubsection}[1]{\subsection{#1}\vspace{-0.05in}}

\newcommand{\tightmypara}[1]{\noindent{\bf {#1}.}}

\newcounter{packednmbr}

\newenvironment{packeditemize}{
\begin{list}{$\bullet$}{
\setlength{\itemsep}{0pt}
\addtolength{\labelwidth}{10pt}
\setlength{\leftmargin}{12pt}
\setlength{\listparindent}{\parindent}
\setlength{\parsep}{2pt}
\setlength{\topsep}{0pt}}}
{\end{list}
}
\newcolumntype{C}[1]{>{\centering\let\newline\\\arraybackslash\hspace{0pt}}m{#1}}

\begin{document}

\newcommand{\sys}{\mbox{{xPlane}}\xspace}
\pagenumbering{gobble}
\pagestyle{plain}

\title{\huge A Roadmap for Enabling a  Future-Proof In-Network Computing Data Plane Ecosystem}

\author{\Large Daehyeok Kim$^\star$, Nikita Lazarev$^\dagger$, Tommy
Tracy$^{\S}$, Farzana Siddique$^{\S}$, Hun Namkung$^\star$ \\James C. Hoe$^\star$, Vyas Sekar$^\star$, 
Kevin Skadron$^{\S}$, Zhiru Zhang$^\dagger$, Srinivasan Seshan$^\star$}

\affiliation{
  \institution{$^\star$Carnegie Mellon University, $^\dagger$Cornell University,  $^{\S}$University of Virginia}
 }
\begin{abstract}
As the vision of in-network computing becomes more mature, we see two parallel
evolutionary trends. First, we see the evolution of richer, more demanding
applications that require capabilities beyond programmable switching ASICs.
Second, we see the evolution of diverse data plane technologies with many other
future capabilities on the horizon. While some point solutions exist to tackle
the intersection of these  trends, we see several ecosystem-level disconnects
today; \eg  the need to refactor applications for new data planes, lack of
systematic guidelines to inform the development of future data plane
capabilities, and lack of holistic runtime frameworks for network operators. In
this paper, we use  simple-yet-instructive  emerging application-data plane
combination to highlight these disconnects. Drawing on these lessons, we sketch
a high-level roadmap and guidelines for the community to tackle these to create
a more thriving  ``future-proof'' data plane ecosystem.
\end{abstract}
\maketitle

\tightsection{Introduction}
\label{sec:intro}
Recent advances in programmable switching ASICs~\cite{www-tofino,
www-xpliant,www-trident4,www-silicon_one} have enabled the network data plane to
move beyond its traditional role of packet forwarding. Today, we have more
sophisticated capabilities to process packets to accelerate both network- and
application-level functions.  This emerging trend toward in-network computing
enables new opportunities to improve performance and lower operational costs of
data center infrastructure~\cite{hotnets17-daiet}.  Indeed, recent efforts have
shown that various datacenter applications, such as key-value
stores~\cite{sosp17-netcache,nsdi18-netchain}, machine
learning~\cite{nsdi21-switchml}, and network functions~\cite{sigcomm17-silkroad,
sigcomm18-sonata} can benefit from in-network computing.

Looking forward, we see that there are increasing demands from applications that
could benefit from in-network computing but cannot be realized on today's
switches. For example, while many recent studies have shown the promise of
sketch-based network monitoring on
switches~\cite{sigcomm20-beaucoup,conext20-fcm}, we observe that it is
infeasible to maintain sketches for multiple sub-populations of interest due to
limited on-chip compute and memory resources. Also, applications that require
inspection of packet payloads, such as an intrusion detection over fragmented
packet streams (\eg TCP streams), or middlebox functions over encrypted packet
streams, are not possible on switches due to the lack of an ability to process
packet payloads. While other devices in the network (\eg a software
switch~\cite{www-vpp,www-dpdk-swx}) could implement some of the necessary
functionality, the design of modern switches makes it difficult to seamlessly
incorporate this external functionality. 



In parallel, we also observe the exciting evolution of data plane technologies
beyond switching ASICs, including FPGA-,  NPU-, and DPU-based networking 
devices~\cite{www-u50,www-cx5,www-netro,www-n3000,www-dpu} that are designed to 
accelerate packets and data processing near the end hosts. Moreover, some switch
vendors even manufacture a switch chassis that is equipped with both switching
ASICs and FPGAs~\cite{www-tahoe-2624}.  While this technology trend seems
promising to support the evolving application demands, as mentioned above,
modern switches have no way to incorporate this external functionality into the
data plane. 

As an illustrative case study at the intersection of these evolutionary trends,
we consider one representative application, multi-dimensional sketch-based
network monitoring and two emerging data plane technologies, FPGA and
PIM-enabled data plane~\cite{www-u50,www-n3000,hpca20-fulcrum,
upmem_hotchips2019, isca21-fimdram-lee}. This case study was just meant as a 
very simple starting point to bootstrap the authors' research collaboration
spanning a cross-disciplinary team of sketch designers, network data plane
researchers, and hardware architects. We explored how we can extend existing
efforts to address this scenario (\eg  leveraging external data plane devices to
address limitations of switch ASICs~\cite{sigcomm20-tea,nsdi21-flightplan}).
Unfortunately, this exercise made it abundantly clear  that there are
fundamental disconnects between the requirements of different participants
involved in this collaboration (\ie sketch developers, hardware architects
networking researcher) and the capabilities of existing solutions. In essence,
we find that most existing efforts are point solutions for particular
combinations of applications and data plane devices  and provide little to no
general capabilities for the relevant participants.

This disconnect between a seemingly simple case study and the capabilities of
existing efforts  naturally suggested  the need for a more principled approach
and design guidelines for creating a thriving future-proof  ecosystem to
accommodate and leverage such evolving applications and data plane
technologies.  This paper is merely a first   step to bootstrap this discussion
in the community. 
  
To this end, we begin by first identifying key stakeholders in the ecosystem
(hardware vendors, application developers, and network operators). We use our
case study lessons to raise essential questions from each of them, and try to
provide guidelines for answering them to derive future-proof approaches.  For
example, hardware vendors need a principled way of choosing data plane
technologies, designing building blocks a chosen data plane, and integrating the
switch ASICs and the data plane given application demands.  Application
developers need a platform  agnostic programming toolchain to avoid re-writing
programs for different platforms.  Network operators should be able to choose
the right hardware platform for their application workloads while being able to
manage resources at runtime.

We also present a preliminary attempt for answering these  questions using our
case study to shed light  on a potential roadmap for creating a more thriving
and future-proof data plane ecosystem. Although our exploration is admittedly
preliminary, we hope that by drawing attention to this ecosystem-level view of
in-network computing, our work opens up the discussion and   new research
opportunities and directions for the community; \eg programming language
support for general applications, runtime resource and state management, and
making the architecture fault tolerant.

\tightsection{Background and Motivation}
\label{sec:motivation}

\tightsubsection{Evolving Application Demands}
\label{sec:motive_app}

Many recent works have shown the promise of in-network computing for various
applications~\cite{sigcomm17-silkroad,osdi20-pegasus,conext20-fcm,sigcomm18-sonata,nsdi21-switchml,sec21-jaqen}
by leveraging programmable switching
ASICs~\cite{www-tofino,www-trident4,www-silicon_one,www-xpliant}.  The key
benefit of running applications on programmable switches is that they are
located at the vantage point in the network where they can naturally observe
packets. While all of these efforts seem promising, we also observe that it is
difficult to support the evolving demands (in terms of workload sizes and
required capabilities) for a wide variety of applications on switches, limiting the potential of in-network
computing.  Here, we introduce three such applications that could benefit from
running inside the network but cannot be supported by the switch ASICs today. 

\tightmypara{Multi-dimensional sketch-based monitoring} Sketching algorithms are attractive as network monitoring capabilities on programmable
switches since they offer high accuracy guarantees and use compact data
structures. A single sketch instance reports statistics (\eg heavy hitters) for
a single sub-population defined by a flow key (\eg IP 5-tuple). We observe that
as the network traffic becomes  diverse~\cite{www-cisco-cloud-report,www-cisco-report} there is an increasing
need for monitoring traffic from multiple dimensions
simultaneously~\cite{sec21-jaqen}. This requires multiple sketch instances 
running on a switch concurrently, which we call multi-dimensional sketching.
However, we find that due to the limited on-chip compute and memory resources and their inefficient allocation, 
the switch can only run a few instances. For example, in our experiments 
with Count-Sketch~\cite{vldb08-countsketch}, we observe that only up to 3 sketch
instances can be implementable on a Tofino switching ASIC (not shown).

\tightmypara{Intrusion detection/prevention systems}
Intrusion detection and prevention systems (IDS/IPS) are another example that
could benefit from running inside the network. Today, they are implemented and
running on x86 servers~\cite{comnet99-bro,www-bro,www-snort} and accelerated
using FPGAs~\cite{osdi20-pigasus}. Typically, they reassemble TCP streams and
perform string matching over reassembled packets to detect malicious traffic.
Running this functionality on switches would reduce the latency overhead of
filtering malicious traffic by avoiding rerouting every packet to servers
running IDS/IPS.  However, it is infeasible to implement such functionalities
due to the inability of switch processing (due to  limited memory and a
constraint computation model) to reassemble packets and inspect payloads.

\tightmypara{Middlebox functions over encrypted traffic}
The majority of today's Internet traffic is encrypted using end-to-end protocols
such as TLS~\cite{dierks1999tls} and its volume keeps
growing~\cite{conext14-https}.  There have been efforts to enable middlebox
functions such as WAN optimization and load balancing over encrypted traffic
on x86 server-based middleboxes~\cite{sigcomm15-mctls,sigcomm15-blindbox}.
However, similar to the intrusion detection example, switches cannot process
encrypted traffic due to their inability to inspect packet payloads and their
limited computation capability.

We argue that the current design of programmable switching ASICs makes it
difficult to meet these evolving application demands. Traditionally, switching 
ASICs are designed to support only stateless or simple stateful functionality,
such as packet forwarding and access control in a synchronous manner at
line-rate (\eg a few Tbps).  Because of this, even today, they are equipped with
a limited amount of fast but expensive memory (\eg SRAM and TCAM) and compute
(\eg simple ALUs and hashing units) on-chip resources, which are infeasible to
extend post-manufacture.  However, many applications, including the above
mentioned ones, may not require synchronous or line-rate processing, but still
could benefit from running inside the network.  For example, multiple sketch
instances can be updated for each incoming packet, asynchronously to packet
routing. However, due to limited on-chip resources, it is impossible to
implement such functionality on switches today. Thus, the mismatch between the
switch data plane design and evolving application demands limits the potential
of in-network computing. 

\tightsubsection{Evolving Data Plane Technologies}
\label{sec:motive_dp}
In recent years, many hardware vendors have manufactured FPGA-,  Netronome's NPU-, and
Nvidia's DPU-based data plane devices~\cite{www-cx5,www-u50,www-n3000,www-netro, www-dpu}
that trade off lower packet processing capacity compared to switch ASICs, for larger memory space and processing capabilities. Also, there are some software-based data planes running on x86 servers~\cite{www-dpdk-swx, nsdi15-ovs, www-vpp}. They can be
programmed by using vendor-supplied APIs~\cite{www-cx5, www-dpu} or high-level
programming languages and libraries like P4~\cite{ccr-p4,www-netcope-p4} and
OpenCL~\cite{www-opencl}. In this section, we introduce the FPGA-based data
plane, which is currently available and already widely used, and the
Processing in memory (PIM)-enabled data plane, which is not mature yet, but seems promising.


\tightmypara{Example 1: FPGA-based data plane}
Recent studies have shown the promises of FPGA-based NICs as efficient
networking devices capable of processing large volumes of data at line
rate~\cite{nsdi18-azure_smartnic, Limago_fpl2019, www-u50}. FPGAs contain an
ample amount of heterogeneous, fine-grain, programmable resources such as
look-up tables (LUTs), DSP slices, distributed on-chip memory, and programmable
I/Os which support high packet processing capacity and flexibility of FPGA-based
NICs.

FPGAs continue to evolve today. In-network processing applications driven by future FPGAs will benefit from a large number of high-bandwidth transceivers (\eg up to twenty 100
GbE interfaces are already available in the Stratix 10
device~\cite{intel_stratix_10}), emerging routing
technologies~\cite{intel_hyperflex} that promise to boost the maximum clock
frequency of the data plane up to 700 MHz, and integrated on-chip systolic
acceleration for high-performance MIMD computing. In addition, today's FPGAs can
be tightly coupled with 3D-stacked high-bandwidth memories (HBM),  which makes
implementation of complex high-throughput applications such as data
sketching~\cite{kulkarni2020hyperloglog} feasible.  Finally, modern FPGAs
achieve very high system integration driven by the 2.5D~\cite{Shuhai_fccm2020}
and through-silicon via (TSV) technologies~\cite{motoyoshi2009through}.  These
allow integrating networking interfaces, data processing units, and memory in a
single package, therefore, reducing the off-chip traffic and further improving
the performance and power efficiency of in-network processing systems.

\tightmypara{Example 2: PIM-enabled data plane}
We observe many in-network computing applications bring data and compute
together~\cite{sigcomm16-univmon, sigcomm18-sonata}.  Seen from this angle,
future data planes that exploit PIM~\cite{pawlowski2011hybrid, jun2017hbm}
appear as a  promising alternative for in-network computing.  PIM addresses the
data access bottleneck due to latency and bandwidth limitations of traditional
memory (\eg DRAM) by performing the computations close to the data, in the
memory unit. And by placing processing elements in every chip, bank, or even
subarray, the resulting parallelism can enable a PIM device to keep up with very
high data rates.  Some PIM products are already on the market, including
UPMEM~\cite{upmem_hotchips2019}, which places a data processing unit at each
bank, and Samsung's FIMDRAM~\cite{isca21-fimdram-lee}, which targets machine
learning applications and integrates a 16-wide SIMD matrix-vector engine within
memory banks, replacing half of the cell array per bank.

\tightsubsection{Motivating Scenario}
\begin{figure}[t!]
\includegraphics[width=0.9\columnwidth]{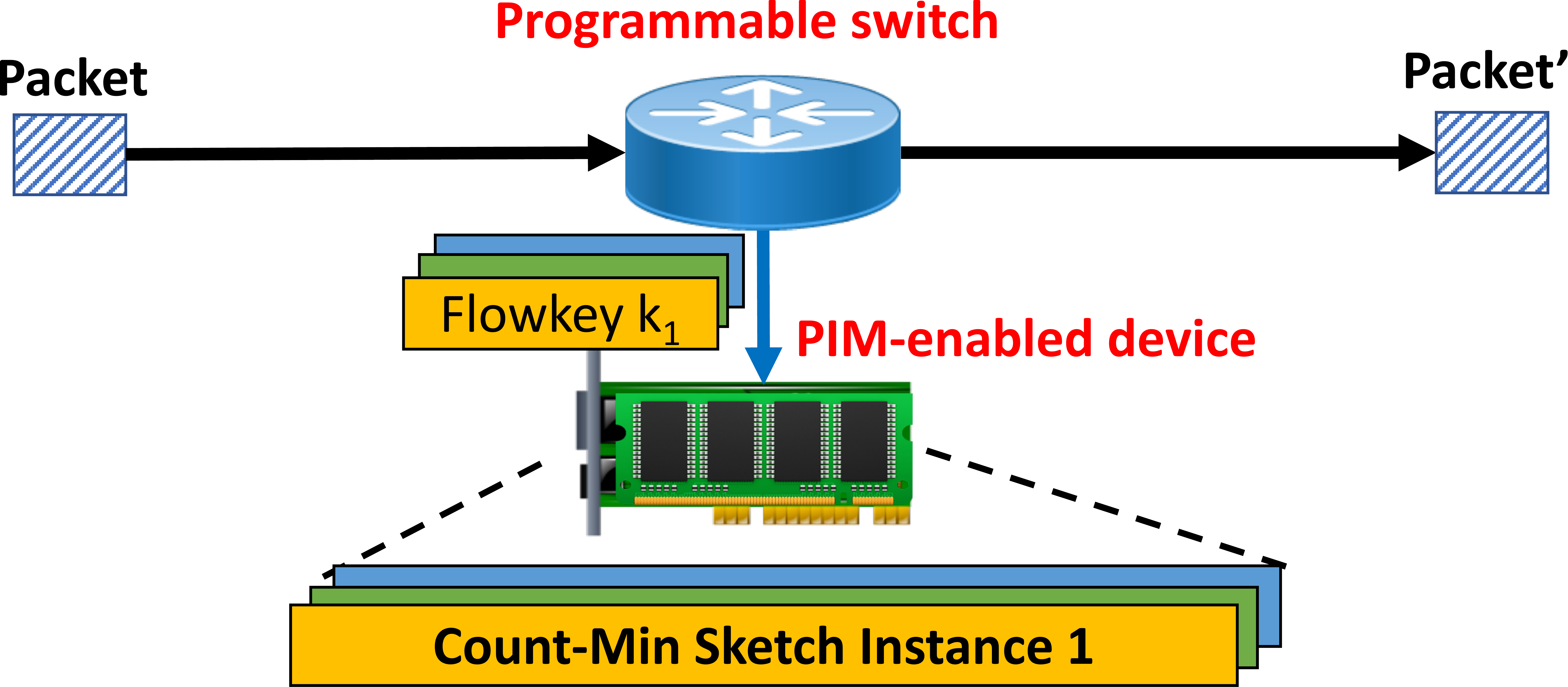}
\caption{Motivating scenario: Running multi-dimensional sketch-based network monitoring on Switch ASIC + PIM-enabled data plane.} 
\label{fig:case_study} 
\vspace{-0.1in}
\end{figure}


Suppose a network operator wants to run multi-dimensional sketch-based network
monitoring on a programmable switch, but due to limited resources of the switch
ASIC, it fails to run. Now, the operator tries to run the application on a
combination of the switch ASIC and PIM-enabled data plane device as it provides
an ample amount of memory resources and associated reconfigurable logic, which
can be complementary to limited resources of the switch ASIC
(\autoref{fig:case_study}). 


At first glance, it appears that   there are some existing efforts in leveraging
external data plane devices that could be potentially applicable, such as: (1)
new  programming languages for heterogeneous data
planes~\cite{sigcomm20-microp4,sigcomm20-lyra}; (2)  frameworks that help to
offload functionality to programmable NICs~\cite{osdi18-floem,sigcomm19-ipipe};
(3)  program partitioning frameworks such as
Flightplan~\cite{nsdi21-flightplan}; and (4) switch resource augmentation
systems such as TEA~\cite{sigcomm20-tea}. 

Unfortunately, as we explore this case study further, we observe that none of
these aforementioned solutions are  directly applicable to this scenario. 
First, programming languages for heterogeneous data planes such as
Lyra~\cite{sigcomm20-lyra} and $\mu$P4~\cite{sigcomm20-microp4} cannot be
applied  because they aim to translate a single program to a single device or a
single type of device (\eg switch ASIC).  Second, programming frameworks that
help to offload functionality to programmable NICs such as
iPipe~\cite{sigcomm19-ipipe} and Floem~\cite{osdi18-floem} do not work for this
scenario because they also consider a case where offloading a single program to
a single target (\ie NIC).  Third, while program partitioning frameworks such as
Flightplan~\cite{nsdi21-flightplan} can statically disaggregates a program to
multiple data plane devices (\eg switches and FPGAs) , they focus on specific
targets (\eg P4-compatible) and do not consider runtime resource management.
Lastly, switch resource augmentation framework such as  TEA~\cite{sigcomm20-tea}
augments switch's resources (\eg SRAM) for state-intensive network functions, it
is unclear whether they applies to other applications with different workload
characteristics (\eg memory access patterns) and other memory technology such as
PIM. 

Moreover, these  above disconnects are not unique to our choice; we observe
that all these existing approaches do not work even for other combinations of an
application and external data plane (in fact, our specific choice of application
and data plane was quite incidental and was actually intended as a hopefully
simple  example to bootstrap a collaborative effort between networking and
computer hardware architects.). This motivates us to think of these disconnects
are perhaps symptomatic of a broader data plane ecosystem-level problem, and
that we need to explore a more principled approach and design guidelines for a
future-proof ecosystem. 
\tightsection{Data Plane Ecosystem: \\ Stakeholders and Requirements}
\label{sec:overview}

Instead of proposing another point solution for a particular combination of an
app and a data plane device, we argue that we need to rethink a future-proof
data plane ecosystem to adapt evolving technologies to support evolving app
demands.  To this end, we begin by identifying the key stakeholders in this
ecosystem and their goals.  There are three key players in the data plane
ecosystem: hardware designers/vendors, application developers, and network
operators.

\tightmypara{Hardware designers/vendors} 
Given application or workload demands (\eg number of traffic flows to monitor
and sketch data structure parameters), hardware vendors need to design data
plane devices (or components) that are connected to switch ASICs while
considering the cost-efficiency, generality (\ie supporting next generation
applications and data plane technology), and extensibility (\eg an ability to
add more memory space or new capabilities). 

\begin{packeditemize}
\item What is the right architecture of external data plane devices for given workloads?
\item What is the right interface between the switch ASIC and external data plane
devices?
\end{packeditemize}

\tightmypara{Application developers}
Given application requirements, application developers need a general way to
write applications (\eg sketch-based monitoring) that can best leverage
resources and capabilities on switch ASICs and other (even future) data plane
devices.

\begin{packeditemize}
\item What is a good programming model that allows for the implementation of a
wide range of applications while hiding the complexity of integrated data
planes?
\end{packeditemize}

\tightmypara{Network operators}
Given application demands, a set of available hardware platforms, and cost
budgets, network operators need to choose  platforms that satisfy their
application requirements while considering generality (\ie supporting future app
demands), extensibility, and cost-efficiency.   

\begin{packeditemize}
\item What hardware should be provisioned for a given application, a set of data
plane devices, and cost budgets?
\item How to manage available data plane resources and route I/O requests to
proper external data plane devices at runtime?
\end{packeditemize}

%

%

\tightsection{Hardware Design Space Exploration}
\label{sec:design_space}

We explore the hardware design space 
 in this section by revisiting  
our case study from~\autoref{sec:motivation} in more detail. We defer questions for other stakeholders to the next section. 

\tightsubsection{Application Workloads}
\label{sec:case_workload}

\tightmypara{Sketch configuration} We assume a network monitoring application
that uses 10 Count-Min Sketch instances to monitor 10 different sub-populations
of traffic defined by 10 different flow keys such as a source IP, destination
IP, and IP 5-tuple. Each instance consists of 3 $\times$ 32-bit counter arrays
of size 64K.  To track heavy flow keys, we assume that the application uses
priority queues. 

\tightmypara{Traffic workload} We assume that the network operator wants to
monitor the Internet backbone traffic. Since the traffic rate in terms of
packets per second is an important factor that can affect the application
performance and hardware design, we analyze a number of publicly available CAIDA
packet traces~\cite{www-caida} and find that an average traffic rate is around
1.43~Mpps. We use this value for the rest of this section. Note that among this
1.43~M packets, around a few hundred of the packets update the heavy flow
priority queue. 

The main goal of hardware designers is to thoroughly understand possible design
options and choose the best one that potentially satisfies network operators and
application developers' requirements. Here, we sketch out one possible
principled way of exploring design spaces.

Hardware designers need to explore three design spaces: (1) Choice of data plane
technology (\eg PIM-enabled data plane), (2) Design of kernels (or
building blocks) for the chosen data plane (\eg a hash table in PIM), and  (3)
Interface between the switch ASIC and the external data plane.

For (1), hardware designers should consider inputs from network operators (\eg
cost budget, performance, and  extensibility) and application developers (\eg
capabilities of interest). Note that in networking context, they need to
consider that the workloads are streaming data. In this case study, we choose
two data plane technologies, PIM and FPGA-enabled data planes, as they are in a
trade-off relationship in terms of performance and extensibility.   

Further, we discuss how designers could explore the design space (2) (in
\autoref{sec:fpga} and \autoref{sec:pim}) and (3) (in \autoref{sec:integration})
for FPGA and PIM-enabled data planes.

\tightsubsection{FPGA-enabled Data Plane}
\label{sec:fpga}

FPGAs extend the model of the traditional programmable switch data planes which
only have a limited support for programmability. The reconfigurable nature of
FPGAs allows fine-granular tuning of both the compute, memory, and I/O
architectures for any given application. This is, in particular, beneficial for
multi-dimensional sketching as its high-bandwidth requirements, the randomness
of memory access patterns, and in-network streaming nature demand specifically
optimized/ customized data paths, memory, and the architecture of processing
elements (PEs). 

Modern FPGAs are capable of running the workloads
such as data sketching at the line rate. The computational part of the
algorithm (\eg hashing engines, comparators) can be implemented with
DSP blocks. For example, a recent study~\cite{kulkarni2020hyperloglog} has shown
that the HyperLogLog sketching algorithm implemented on an FPGA with high-level
synthesis tools consumes up to 16.4\% of available DSPs on an Xilinx
XCVU9P device~\cite{xilinx_vertex_XCVU9P} when targeting the throughput of 9.35
GB/s, which is substantially more than the required bandwidth of $0.51$~GB/s in
our case study.  Similarly, modern FPGAs contain up to $350$~Mb of on-chip
memory which is sufficient to accommodate both the sketching counters
($61.4$~Mb) and key priority queues ($12.8$~Kb). 

Deployment of in-switch algorithms on an FPGA comes with certain challenges and
design decisions. One of them is how to cleanly decouple computation, temporal
scheduling, and data placement in the design specification of the FPGA
accelerator~\cite{heterocl_FPGA_2019}. Each of these components has its own
trade-offs depending on the required performance and the area/cost budget. For
example, hash functions can be implemented using general-purpose LUTs, or DSPs,
or even MIMD tensor cores available in modern FPGAs. The algorithm's data can be
placed in a large variety of possible configurations of distributed on-chip
memory and even in the external HBM if an algorithm with a large memory
footprint is concerned. These challenges are similar to the traditional workflow
of a hardware accelerator design, however in-network line-rate performance
requirements of in-switch computing add more constraints in the design space.


%

%
%
%
%

\tightsubsection{PIM-enabled Data Plane}
\label{sec:pim}

PIM~\cite{pawlowski2011hybrid, jun2017hbm} enables computation to be performed
in or near memory~\cite{lenjani2020fulcrum, hsieh2016accelerating}, minimizing
the overhead of data movement between processing units and memory. Consequently,
PIM is a potential data plane candidate that provides the high bandwidth
required for processing multi-dimensional sketches at line-rate. This bandwidth
is critical for the fetch-modify-write operations needed for counter updates. 

Sketching requires considerable compute throughput, especially to perform the
hash computations to index  memory arrays for each counter operation. If we
consider the sketch configuration and workload of \autoref{sec:case_workload},
we need to update $3\times10\times1.43 = 42.9$~M counters per second. Each of
these updates consists of a load, an increment, and a store. Here load and store
are memory operations, and increment is an integer compute operation. As a
result, the data plane needs to have at least $42.9\times4\times2 = 342.2$~MB/s
memory bandwidth to support $42.9$M updates per second. 


Considering the bandwidth requirement and the poor locality of the sketches, we
argue that bank-level PIM processing would be suitable for our purpose. The
reason being, multiple counters could span across multiple banks, and the logic
layer processing element works on one bank at a time. As a result, implementing
sketch processing on the bank would allow us to perform concurrent counter
updates, leveraging the bank-level parallelism.

To track top-100 heavy flow keys, we use a heap-based priority queue. The memory
capacity requirement for this is $100 \times 4 \times 2 \times 2 = 1600$~B.
Considering the small memory footprint, and to avoid memory accesses for this
small structure, we hypothesize that a dedicated scratchpad for the priority
queue on the memory logic layer would suffice to achieve our $1.43 \times 4 =
5.72$~MB/s bandwidth. 

\tightsubsection{Switch--External Data Plane Integration}
\label{sec:integration}

\begin{table}[t!]
\footnotesize
\begin{tabular}{@{}lp{1.3cm}p{3cm}l@{}}
		& {\bf On-chip}     & {\bf Off-chip, On-chassis}                  & {\bf Off-chassis} \\ \midrule
{\bf Interface}       & 3D stack &  High BW interface, PCIe & Ethernet    \\ \midrule
{\bf Performance}     & High        & Medium                                & Low         \\ \midrule
{\bf Extensibility}   & Low         & Medium                                & High        \\ \bottomrule
\end{tabular}
\caption{Comparison of design options for a switch ASIC--external data plane integrated architecture.}
\label{tbl:design_space}
\vspace{-0.3in}
\end{table}
After choosing a data plane option (\ie PIM or FPGA-based), hardware designers
now need to consider how to integrate the switch ASIC and data plane component.
We find that there are three design options depending on how they are
interconnected, and the options present a trade-off between performance (\ie
throughput and latency) and extensibility. 
\autoref{tbl:design_space} summarizes these options. 

\tightmypara{\bf Option 1: On-chip} A PIM or FPGA-enabled data plane component is located on the switching ASIC chip. 
This can be implemented with 3D stacking technology~\cite{zhang2014top, de2018design}, which allows for stacking dies of different capabilities (\eg FPGA or PIM). With the ASIC and data plane component closely interconnected, on-chip integration can provide high bandwidth (\eg 160~GBps~\cite{zhang2014top}) and low access latency (\eg 12~$ps$~\cite{zhang2014top}).  
However, such high performance comes at the cost of lower extensibility; the type of reconfigurable logic and capacity are fixed at manufacturing time.

\tightmypara{\bf Option 2: Off-chip, On-chassis} A data plane component is located outside the switching ASIC, but on the same chassis. They are connected through either a custom high bandwidth interconnect technology (\eg 2.5D silicon interposer~\cite{wang2014high}) or existing technology such as PCIe. This option could achieve lower performance than the above option, but provide higher extensibility by allowing for plugging in new components.

\tightmypara{\bf Option 3: Off-chassis} A data plane component is located outside the
switch chassis and connected to the switching ASIC through a network link (\eg
via Ethernet). This option provides the highest extensibility as network
operators can add or remove any type of component outside the chassis. In terms
of performance, it achieves the lowest throughput and the highest latency
compared to the other options as data is transmitted through the network link.
Some existing work such as TEA~\cite{sigcomm20-tea} and GEM~\cite{hotnets18-gem}
already demonstrates the feasibility of this approach for a limited set of
applications.

\begin{figure}[t!]
\includegraphics[width=\columnwidth]{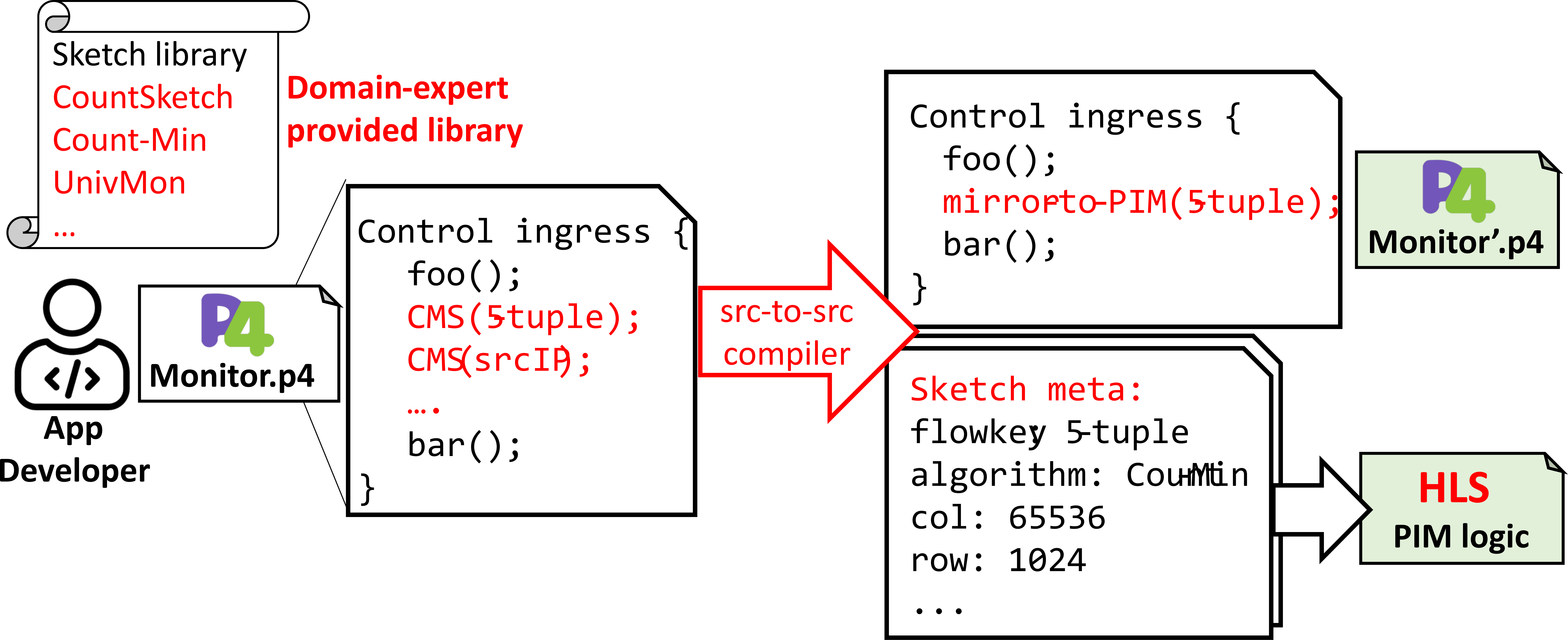}
\vspace{-0.25in}
\caption{Application development workflow.} 
\label{fig:programming} 
\end{figure}

\tightsection{Other Stakeholders and Open Questions}

\tightsubsection{Application Developers}
Ideally, application developers should be able to
implement an application (\eg multi-dimensional sketch-based monitoring) and use it for different platforms. One approach is providing platform-agnostic libraries of capabilities required by the application (\eg a library of sketch algorithms).

\autoref{fig:programming} illustrates an example of such a programming model. 
The app developer writes a program with a library of sketch algorithms
provided by domain experts (or library developers). In this example, we assume
that the switch ASIC supports P4 as a programming language, and the PIM
component supports some high-level synthesis language (\eg C++ with
OpenCL~\cite{www-opencl}). And the sketch algorithms are exposed as P4 APIs (\eg
as P4 control blocks).  Once the developer writes a P4 program (monitor.p4), a
src-to-src compiler translates it into that will be loaded to the switch
ASIC (monitor$^\prime$.p4) and a description of metadata of sketch instances
that will be used as an input for target-specific src-to-src compiler that
generates a PIM program (\eg in C++). 

\begin{figure}[t!]
\includegraphics[width=\columnwidth]{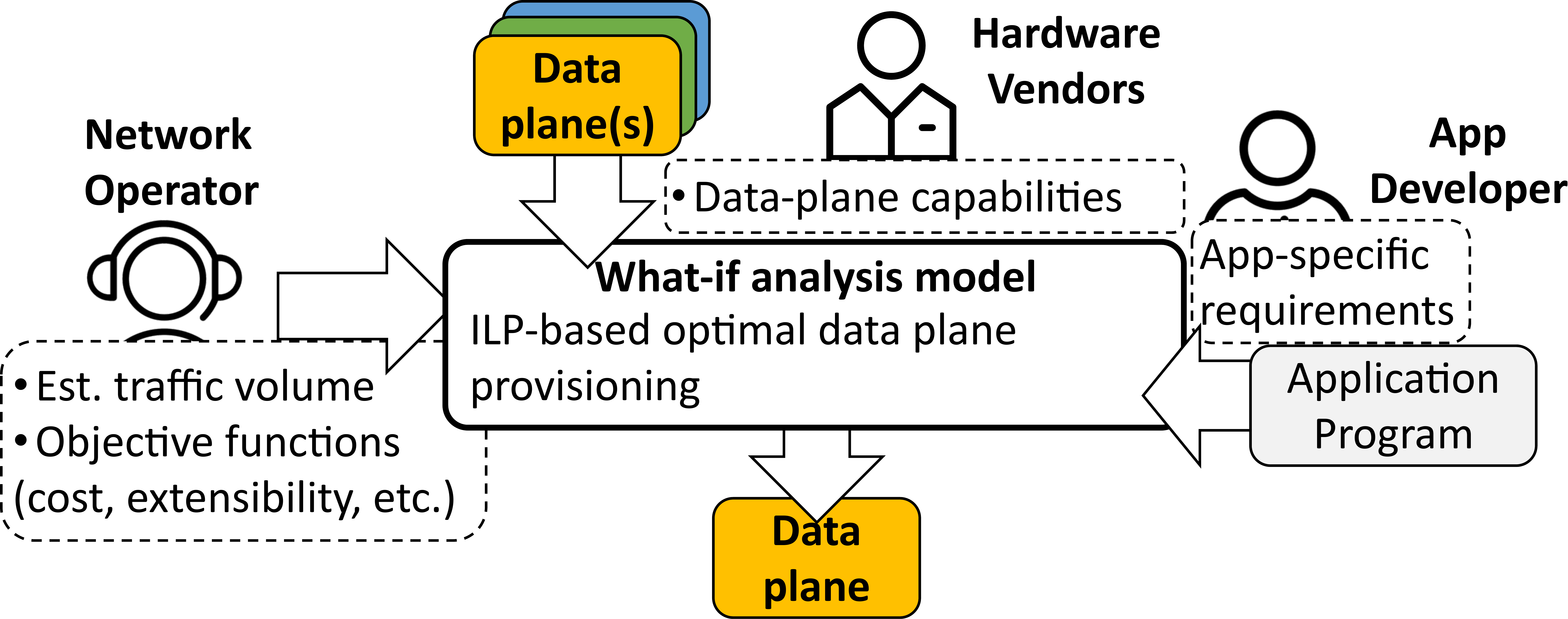}
\vspace{-0.25in}
\caption{What-if analysis for data plane provisioning.} 
\label{fig:whatif} 
\vspace{-0.15in}
\end{figure}

\tightsubsection{Network Operators}
Given a set of data plane options from hardware vendors and apps 
from app developers, network operators now have to provision the right external
data plane. Key decision factors would be application demands, extensibility,
generality, and cost budget. Network operators should be able to find an optimal
data plane  in terms of the above factors.

Here, we outline one potential way of finding an (near-) optimal solution based
on a ``what-if'' analysis.\footnote{We omit the details due to space constraints. The details 
can be found in our technical report~\cite{whatif-tr}.} \autoref{fig:whatif} illustrates an example framework
that takes inputs from all three stakeholders and tries to find an optimal
solution for the data plane provisioning problem.  Specifically, it models the
problem as an integer linear program (ILP) and processes inputs to plug them
into the ILP-based model.  For example, a Network operator provide an estimated
traffic volume (\eg in bps or pps) and their objective functions (\eg optimize
the capital cost), and an application developer provides a application program
and its app-specific requirement (\eg need a particular sketching algorithm).
Hardware vendors provide  data plane platforms and their capabilities. Then the
what-if analysis framework finds an optimal data plane platform based on an
objective function.

\tightsubsection{ Open Questions}
\label{sec:discuss}

We conclude by highlighting a subset of open questions and
future research directions by each stakeholder.

\tightmypara{Hardware designers and vendors}
In previous section, we discuss one possible  way of designing capabilities required by a sketch-based application on PIM and FPGA-based data plane. An open question is whether there is a more principled and data plane agnostic way of exploring the design space for designing capabilities. 

\tightmypara{Application developers} While we illustrate a platform-agnostic library-based
and parameterized programming model for external data plane devices as a potential approach for the sketch-based monitoring, it would be interesting to see if 
this model can be generally applicable to other applications. 
Another potential direction would be to design a new programming language that is agnostic to data plane platforms.

\tightmypara{Network operators}
In an integrated data plane platform, there are different types of resources and capabilities spread across multiple data plane devices. We need a run-time environment support for network operators to manage available resources and allocate workload to devices properly. 
Also, a failure of any data plane devices in the platform can affect the correctness of applications running on it. In particular for stateful  applications (\ie losing state due to failure can affect the correctness or behavior of the app), making them tolerant to failure is not trivial.

{\bibliographystyle{ACM-Reference-Format}
\bibliography{ref}}

\end{document}